\documentclass{phbauth}
\usepackage{graphicx}

\begin{document}

\begin{frontmatter}

\title{Transport and magnetic properties of La$_{1-x}$Ca$_x$MnO$_3$-films ($0.1<x<0.9$)}


\author{Gerhard Jakob\thanksref{thank1}},
\author{Frank Martin},
\author{Stefan Friedrich},
\author{Wilhelm Westerburg},
\author{Markus Maier}
\address{Institute of Physics, Johannes Gutenberg-University, 55099 Mainz, Germany}

\thanks[thank1]{Corresponding author. Present address: Institute of
Physics, Johannes Gutenberg-University, Staudinger Weg 7, 55099 Mainz, 
Germany. E-mail: \makebox{jakob@mail.uni-mainz.de}}

\begin{abstract}
By laser ablation we prepared thin films of the colossal magnetoresistive compound
La$_{1-x}$Ca$_x$MnO$_3$ with doping levels $0.1<x<0.9$ on MgO substrates. X-ray diffraction revealed epitaxial
growth and a systematic decrease of the lattice constants with doping. The variation of the transport and magnetic properties in this
doping series was investigated by SQUID magnetization and electrical transport measurements. 
For the nonmetallic samples resistances up to $10^{13}\Omega$ have
been measured with an electrometer setup. While the transport data indicate polaronic transport for the 
metallic samples above the Curie temperature the low doped ferromagnetic insulating samples show 
a variable range hopping like transport at low temperature.
\end{abstract}

\begin{keyword}
colossal magnetoresistiviy; thin films; La$_{1-x}$Ca$_x$MnO$_3$;
\end{keyword}
\end{frontmatter}

\section{Introduction}

The phase diagram of La$_{1-x}$Ca$_x$MnO$_3$ shows 
different ferromagnetic and antiferromagnetic phases as well as
insulating and metallic phases \cite{Schiffer95} and huge 
magnetoresistive effects have been discovered in thin films of 
manganites \cite{Helmolt93} near doping levels $x=0.3$, where
the ferromagnetic Curie temperature and conductivity are highest.
We investigated the film growth and magnetotransport properties 
of thin films with doping levels 
$x=0.08, 0.125, 0.18, 0.25, 0.33, 0.4, 0.5, 0.6, 0.75, 0.85$. 

\section{Results and discussion}

The thin films were laserablated onto MgO substrates. Due to the rather large lattice mismatch
the films will show a high number of defects at the substrate film
interface and epitaxial strain is released at the early stages of film growth.
Thus for our 270 nm thick films the doping effects should not be obscured by lattice strain. 
In X-ray diffraction only (00$\ell$) reflections could be observed for
films with doping levels $x\le 0.5$. For higher $x$ additionally a 
competing ($0\ell\ell$) oriented growth was visible. The rocking curve widths $\Delta\omega$
were in the range $0.1^\circ<\Delta\omega<  0.6^\circ$. 
\begin{figure}[htbp]
\begin{center}\leavevmode
\includegraphics[width=0.8\linewidth]{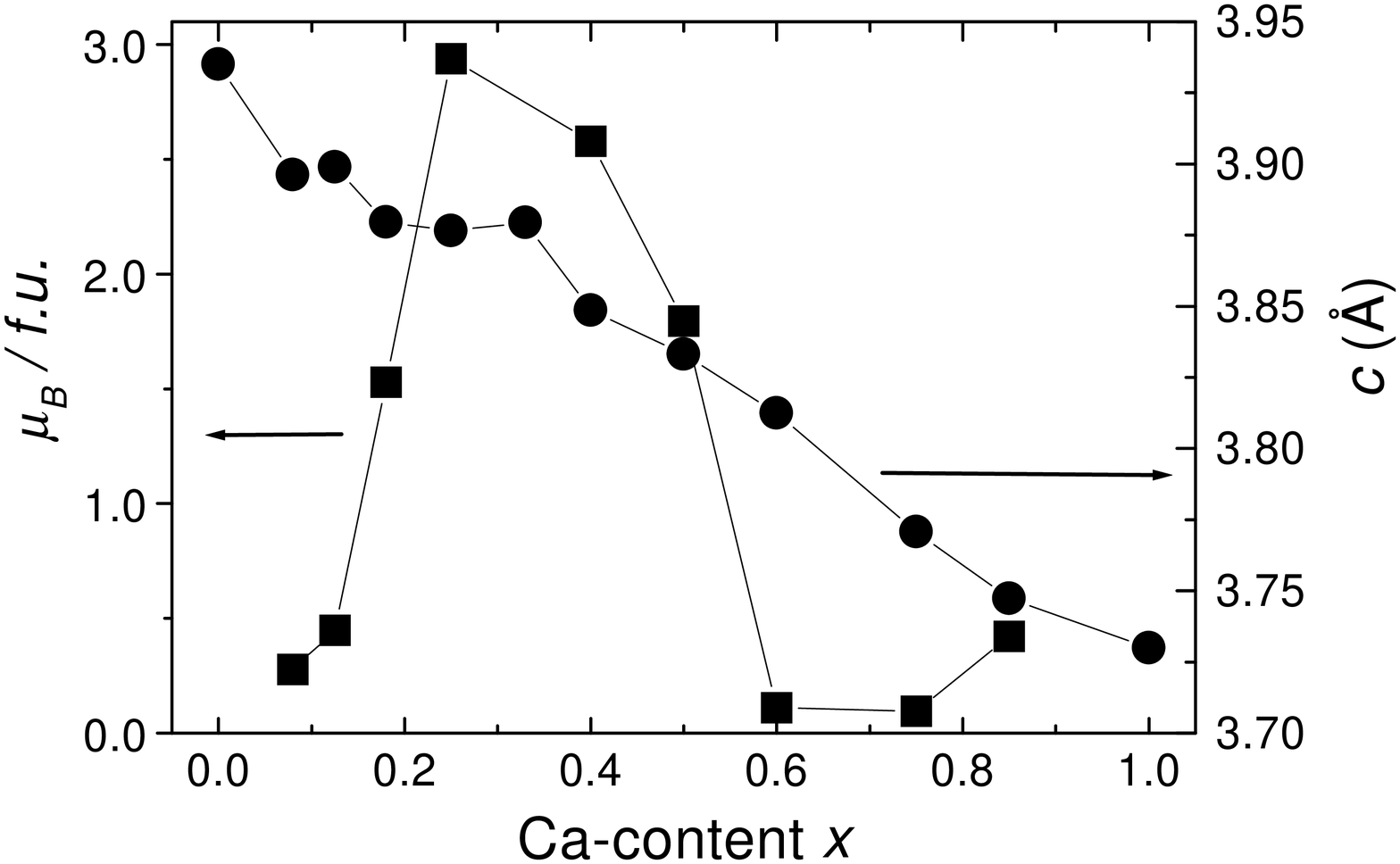}
\caption{ 
Lattice constants and spontaneous magnetic moment as function of doping level $x$.}
\label{AundMvonX}\end{center}\end{figure}
The systematic decrease of the lattice 
constants with doping shown in Fig.~1 indicates a stoichiometric transfer from the target 
to the film. The values for the undoped compounds are taken from literature
\cite{Wollan55}. 
Scanning electron microscopy and scanning force microscopy showed for
$x<0.4$ island growth while for higher doping rectangular platelike 
structures were seen.
Measurements of the temperature dependence of the spontaneous magnetisation in a magnetic 
field of 50 mT with a SQUID magnetometer, showed that the ferromagnetic metallic region of 
the phase diagram in the film coincides with that of bulk material \cite{Schiffer95}.
The low temperature values normalized per formula unit (f.u.) are displayed in Fig.~1.
Above $x>0.5$ the vanishing magnetic hysteresis is in agreement with an antiferromagnetic 
insulating state. However, the sample with $x=0.85$ shows again a ferromagnetic hysteresis loop.

The low temperature coefficient of the resistivity is positive only for films
with $0.2\le x\le 0.5$ as is shown in Fig. 2. Above the Curie temperature samples in this
regime showed polaronic conduction \cite{Jakob98_2}. Outside this doping range the 
temperature coefficient is always negative. For clarity we plotted in Fig.~2 only a selection 
of the curves revealing the metal-insulator transition in this doped films.
The metallic samples have beeen measured by a standard four point technique.
The huge changes in resistance for films outside the metallic regime required the use 
of an electrometer circuit for measurements above 100 M$\Omega$. 
\begin{figure}[htbp]
\begin{center}\leavevmode
\includegraphics[width=0.8\linewidth]{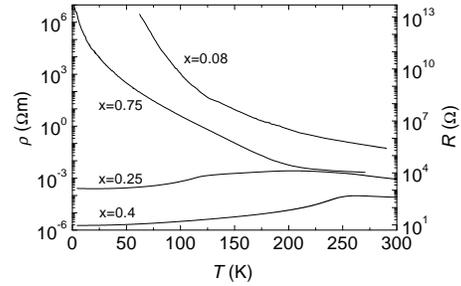}
\caption{ 
$\rho(T)$ for the doping levels given in the figure.}
\label{RvonTundx}\end{center}\end{figure}
Leakage currents were minimised
by use of guarded teflon insulated triaxial cables allowing measurements up to 10 T$\Omega$.  
A pronounced negative magnetoresistance was observed for all the insulating samples 
below the magnetic phase transition temperaures with $R(B=0)/R(B=8{\rm T})\approx 10$.
However, the transport mechanism is different for low ($x<0.2$) and high ($x>0.6$) doped 
samples. On the low doped ferromagnetic side of the phase diagram a variable range hopping 
(VRH) transport describes the measured data well as is obvious from Fig.~3. 
Application of a magnetic field increases the hopping probability without qualitatively
changing the temperature dependence $\rho(T)$. This is different for the Ca-rich samples
where a VRH like transport cannot be identified. 
\begin{figure}[htbp]
\begin{center}\leavevmode
\includegraphics[width=0.8\linewidth]{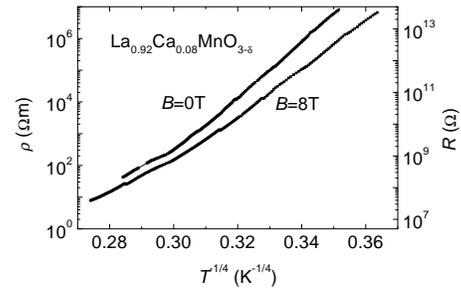}
\caption{ 
$\rho(T)$ plotted versus $1/T^{0.25}$.}
\label{RVRH}\end{center}\end{figure}
\vspace*{-0.8cm}

\begin{ack}
This work was supported by the Deutsche Forschungsgemeinschaft through Project No: JA821/1-3.
\end{ack}


\begin{thebibliography}{9}
\bibitem{Schiffer95} P. Schiffer et. al., Phys. Rev. Lett. {\bf 75}, 3336 (1995).
\bibitem{Helmolt93} R. von Helmolt et. al., Phys. Rev. Lett. {\bf 71}, 2331 (1993).
\bibitem{Wollan55} E.O. Wollan {\it et al.}, Phys. Rev. {\bf 100}, 545 (1955).
\bibitem{Jakob98_2}G.~Jakob {\it et al.}, Phys.\ Rev.\ B {\bf 58}, 14966 (1998).


\end{thebibliography}
\end{document}